\documentclass{article}

\usepackage{epsf}
\begin{document}
\title{Surface effects in the crystallization process\break{} of elastic flexible polymers}
\author{Stefan Schnabel\thanks{Corresponding author: schnabel@itp.uni-leipzig.de}, Thomas Vogel, Michael Bachmann,\\
and Wolfhard Janke\thanks{\{vogel,bachmann,janke\}@itp.uni-leipzig.de; http://www.physik.uni-leipzig.de/CQT.html}\\[.5cm]
{\normalsize{Institut f\"ur Theoretische Physik and Centre for Theoretical Sciences (NTZ),}}\\
{\normalsize{Universit\"at Leipzig, Postfach 100920, 04009 Leipzig, Germany}}}

\date{}

\maketitle

\begin{abstract}
\noindent
Investigating thermodynamic properties of liquid-solid transitions of
flexible homopolymers with elastic bonds by means of multicanonical
Monte Carlo simulations, we find crystalline conformations that
resemble ground-state structures of Lennard-Jones clusters. This
allows us to set up a structural classification scheme for
finite-length flexible polymers and their freezing mechanism in
analogy to atomic cluster
formation. Crystals of polymers with ``magic length'' turn out to be perfectly icosahedral.
\end{abstract}

\section{Introduction}
Small crystals of cold atoms such as argon~\cite{wales1} and 
spherical virus hulls enclosing the
coaxially wound genetic material~\cite{chiu1, busta1} exhibit it, and -- as we will show 
here -- also elastic
flexible polymers in the solid state: an icosahedral or icosahedral-like shape. 
But why is just the icosahedral assembly naturally favored? The reason is that the 
arrangement of a finite number of constituents (atoms, proteins, monomers) 
on the facets of an icosahedron optimizes the interior space filling and thus reduces energy. 

\begin{figure}
\centerline{\epsfxsize=6cm \epsfbox{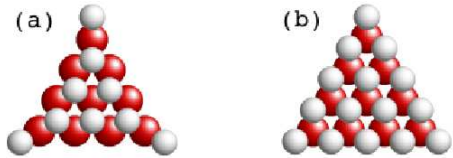}}
\caption{\label{fig:mam} 
(a) Anti-Mackay and (b) Mackay growth overlayer on the facet
of an icosahedron~\cite{northby1}. 
}
\end{figure}

Many-particle systems governed by van der Waals forces
are typically described by Lennard-Jones (LJ)
pair potentials
\begin{displaymath}
E_{\rm LJ}(r_{ij})=4\epsilon[(\sigma/r_{ij})^{12}-(\sigma/r_{ij})^{6}],
\end{displaymath}
where $r_{ij}$ is the distance between two atoms located at ${\bf r}_i$ and ${\bf r}_j$ ($i,j=1,\ldots,N$), 
respectively. An important example are atomic clusters whose structural properties
have been subject of numerous studies, mainly focusing on
the identification of ground-states and their classification. It has 
been estimated~\cite{doye1} that icosahedral-like LJ clusters are favored for systems with $N<1690$
atoms. Larger systems prefer decahedral structures until for $N>213000$ face-centered cubic (fcc)
crystals dominate. In the small-cluster regime, energetically optimal icosahedral-like structures 
form by atomic assembly in overlayers on facets of an icosahedral core. There are two generic
scenarios (see Fig.~\ref{fig:mam}):
Either hexagonal closest packing (hcp) is energetically preferred (anti-Mackay growth), or
the atoms in the surface layer continue the fcc-shaped tetrahedral segment of the interior
icosahedron (Mackay growth)~\cite{northby1}.
The surprisingly strong dependence of structural
liquid-solid transitions on the system size has its origin in the different structure 
optimization strategies. ``Magic'' system sizes allow for the formation of most stable
complete icosahedra ($N=13,55,147,309,561,923$)~\cite{doye2}. Except for a few exceptional 
cases---for $13\le N\le 147$ these are $N=38$ (fcc truncated octahedron), $75$-$77$ (Marks decahedra),
$98$ (Leary tetrahedron), and $102$-$104$ (Marks decahedra)~\cite{doye1}---LJ clusters 
typically possess an icosahedral core and overlayers are of 
Mackay ($N=31$-$54,82$-$84,86$-$97,99$-$101,105$-$146,\ldots$) or anti-Mackay 
($N=14$-$30,56$-$81,85,\ldots$) type~\cite{frant1}.

Although it seems that there are strong analogies in liquid-solid transitions of LJ clusters and 
classes of flexible polymers, comparatively few systematic attempts were undertaken to
relate for finite systems structural properties of LJ clusters and frozen polymers~\cite{karp1,doye3}. 
In contrast, the analogy of the generic phase diagram for colloids and polymers has been addressed
in numerous studies (for a recent overview, see, e.g., Ref.~\cite{binder1}).

Here, we systematically analyze for a frequently used 
flexible LJ polymer model with anharmonic springs~\cite{FENE,binder3} the 
formation of frozen conformations in conformational liquid-solid transitions
and find that for the chain lengths studied ($N\le 309$) the ground-state conformations resemble 
structures known from LJ clusters of corresponding size.
Our results for the peak structures of energetic and structural fluctuating quantities and the 
interpretation of the liquid-solid transitions invalidate recent results from studies of the same 
model~\cite{parsons1}, but are consistent with
former studies of homopolymers with harmonic bonds~\cite{doye3} and analyses of LJ cluster 
formations~\cite{frant1}. In analogy to a recent analysis of the crystallization
of flexible lattice polymers~\cite{vbj1} and LJ cluster studies~\cite{frant1}, there is no
clear size-dependent scaling behavior for energetic and structural fluctuations near the 
liquid-solid transition in the icosahedral regime being in the focus of our study. As expected,
liquid-solid and $\Theta$ collapse transition seem to remain separate transitions in the 
thermodynamic limit~\cite{binder1,vbj1,binder2}. 

\section{Model and method}
For our study, we employ a polymer model with truncated and
shifted pairwise LJ potential,
\begin{displaymath}
E_{\rm LJ}^{\rm mod}(r_{ij})=E_{\rm LJ}(\min(r_{ij},r_c))-E_{\rm LJ}(r_c),
\end{displaymath}
and finitely extensible nonlinear elastic (FENE) anharmonic bonds~\cite{FENE,binder3} between
adjacent monomers,
\begin{displaymath}
E_{\rm FENE}(r_{i\,i+1})=-KR^2\ln\{1-[(r_{i\,i+1}-r_0)/R]^2\}^{1/2}.
\end{displaymath}
For the parametrization we follow Ref.~\cite{binder3}. In our units, the LJ parameters are set to 
$\epsilon=1$ and $\sigma=2^{-1/6}r_0$ with the potential minimum at $r_0=0.7$ and the cutoff at 
$r_c=2.5\sigma$. The FENE potential has a minimum coinciding with $r_0$ and it diverges for 
$r\rightarrow r_0\pm R$ with $R=0.3$. The spring constant $K$ is set to $40$.
Finally, the total energy of a polymer conformation ${\bf X}=({\bf r}_1,\ldots,{\bf r}_N)$ is
given by
\begin{displaymath}
\label{eq:poly}
E({\bf X})=\frac{1}{2}\sum_{i,j=1\atop i\neq j}^NE^{\rm mod}_{\rm LJ}(r_{ij})+\sum_{i=1}^{N-1}E_{\rm FENE}(r_{ii+1}).
\end{displaymath}
In order to sample the entire state space with high accuracy, we performed multicanonical 
Monte Carlo simulations~\cite{muca1,muca1b}, 
where recursively calculated multicanonical weight factors~\cite{muca2,muca2b} 
deform the energy histogram in such a way that entropically suppressed low-energy conformations are 
sampled as frequent as high-energy states which are of less interest for the conformational
transitions discussed here.
In the final production runs, at least $10^{10}$ updates were generated, 90\% of which were 
simple displacements of single monomers and 10\% bond exchange updates altering monomer linkages. 
All conformations we refer to as 
ground states in the following were identified in these multicanonical simulations
and refined with standard minimizers.

\section{Results and discussion}
Before discussing size-dependent thermodynamic properties of structure formation in 
liquid-solid transitions of the flexible FENE polymers, we discuss the structural composition
of ground-state conformations in the icosahedral regime. Since the FENE bond potential enables
almost nonenergetic deformations of covalent bonds near the LJ minimum,
polymer ground-state morphologies resemble corresponding shapes of LJ clusters. It should be noted, 
however, that the covalent bonds explicitly
break rotational symmetries and, in consequence, the polymer ground states are highly degenerate
and metastable. 

\begin{figure}
\centerline{\epsfxsize=7cm \epsfbox{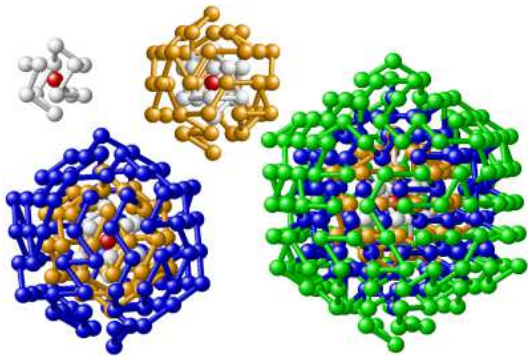}}
\caption{\label{fig:ico} 
``Magic'', i.e., perfectly icosahedral ground-state conformations 
of elastic flexible polymers with 13, 55, 147, and 309 monomers, as found in the 
multicanonical simulations. The characteristic Mackay shells are shaded differently.
}
\end{figure}

At zero temperature, similar to LJ clusters, also flexible FENE polymers with $N\ge 13$
monomers usually build up structures with icosahedral 
cores. The smallest icosahedral structure can be formed by the 13mer: A central monomer
is surrounded by twelve nearest neighbors. For larger system sizes, additional monomers form 
new shells around the central 13mer and the next complete icosahedron with $N=55$ contains two 
completely filled shells---and represents the core of the triple-layer icosahedron 
with $N=147$ etc. Corresponding ground-state conformations with almost 
perfectly icosahedral shape as found in our multicanonical simulations
are shown in Fig.~\ref{fig:ico}.
In complete analogy to the LJ clusters, 
ground-state morphologies of FENE polymers with intermediate chain lengths ``grow'' 
by forming anti-Mackay or Mackay overlayers, except for special cases such as
the polymer with $N=38$. If only a few excess monomers remain after compact core formation, 
these are bound to the facets of the icosahedral core in the optimal distance, thus
filling the available space at the facet most efficiently by forming an anti-Mackay layer 
[Fig.~\ref{fig:mam}(a)]. For larger systems, 
a crossover to the formation of Mackay-type layers 
[Fig.~\ref{fig:mam}(b)] occurs as it allows to place additional monomers on the 
edges. This increase in structural compactness energetically overcompensates the occurrence of partly
nonoptimal distances which induces strain into the structure (this is the reason why for very
large systems fcc crystals are favored~\cite{doye1}). 

\begin{figure}
\centerline{\epsfxsize=7.5cm \epsfbox{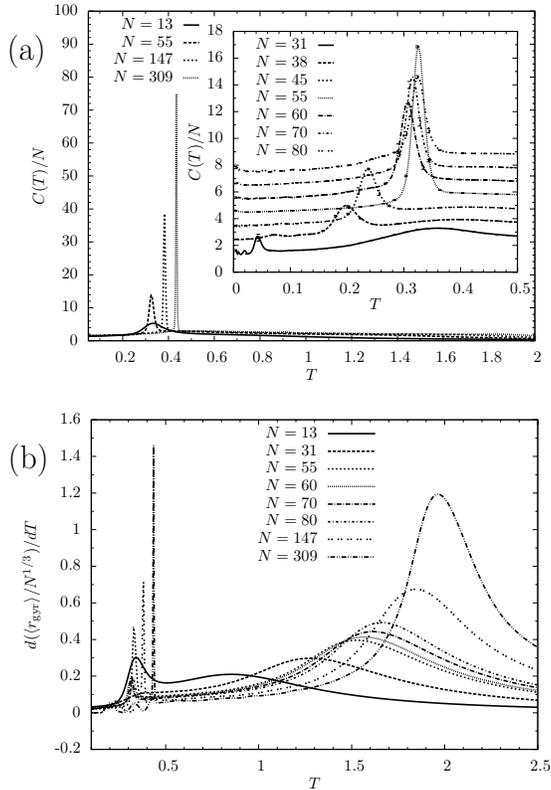}}
\caption{\label{fig:flucts} 
(a) Specific heats (in the inset shifted by a constant value for the sake of clarity) and 
(b) fluctuations of the mean radius of
gyration $d\langle r_{\rm gyr}\rangle/dT$ as functions of the temperature for chains of different length.
Jackknife~\cite{jack1,jack2} error bars are as small as the line thickness.
}
\end{figure}

The liquid-solid transition of the rather short polymers considered here 
is not a phase transition in a strict thermodynamic sense and its strength depends on
the chain length. Figure~\ref{fig:flucts}(a) shows the specific heat
$C_V=(\langle E^2\rangle-\langle E\rangle^2)/T^2$ as a function of the temperature $T$ for several
exemplified polymers with different chain lengths, and in Fig.~\ref{fig:flucts}(b), the fluctuations
$d\langle r_{\rm gyr}\rangle/dT$ of the radius of gyration 
$r_{\rm gyr}=\sqrt{\sum_i ({\bf r}_i-\bar{\bf r})^2/N}$ (with $\bar{\bf r}=\sum_i {\bf r}_i/N$)
are plotted. The latter curves exhibit noticeable fluctuations in the high-temperature regime which
indicate the collapse transition from extended random coils to more compact, globular conformations
(the ``liquid'' state). Energetic fluctuations are rather weak such that this transition can hardly
be identified in the specific-heat curves. Much more striking is the sharpness of the 
low-temperature peaks exhibiting clear signals for a further structural compactification
[Fig.~\ref{fig:flucts}(b)] and energetic optimization [Fig.~\ref{fig:flucts}(a)]---the liquid-solid
transition\footnote{Our results are in correspondence with studies of LJ clusters~\cite{frant1} and polymers with 
harmonic bonds~\cite{doye3}. However, there is significant mismatch with the former FENE polymer 
melting studies of Ref.~\cite{parsons1,parsons2}, where the energy space was artificially restricted to
$E\in [-4N,0)$, thus neglecting the lowest-energy states, which are, however, 
statistically relevant for the thermodynamic analysis of the liquid-solid transition.
}.

There are three characteristic features: First, the transitions 
are particularly strong for chains with ``magic'' length, which possess perfectly
icosahedral ground-state
morphology (e.g., $N=13,55,147,309$). A second type of liquid-solid
transition consists of two steps, at higher temperatures the formation of an icosahedral core 
with anti-Mackay overlayer that transforms at lower temperatures 
by monomer rearrangement at the surface into
an energetically more favored Mackay layer (``solid-solid'' transition). 
This is the preferred scenario for most of the chains 
with lengths in the intervals $31\le N\le 54$ or $81\le N\le 146$ 
that make the occupation of edge positions in the outer shell 
unavoidable. In most of the remaining cases, typically anti-Mackay layers
form. 
To conclude, although the elastic polymers are entropically restricted by the covalent bonds,
we find that the general, qualitative behavior in the freezing regime exhibits noticeable similarities
compared to LJ cluster formation.

\begin{figure}
\centerline{\epsfxsize=8cm \epsfbox{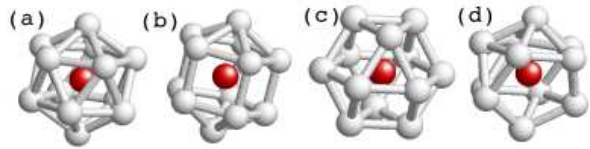}}
\caption{\label{fig:shells} 
Smallest single-shell cores of compact conformations:
(a) icosahedron, (b) elongated pentagonal pyramid, (c) cuboctahedron (fcc), 
and (d) incomplete icosahedron. Sticks represent nearest-neighbor contacts in the shell, not bonds.
}
\end{figure}

For the quantitative analysis of the structuring process we introduce an 
``order parameter'' enabling the discrimination of conformational phases in the different 
scenarios. At low temperatures, all monomers in the core are always surrounded by 12 
almost equidistant nearest neighbors. 
However, the arrangement of these neighbors can differ. The lowest-energy assembly 
is the smallest icosahedral cell with 30 nearest-neighbor 
contacts in the shell [Fig.~\ref{fig:shells}(a)]. All icosahedral conformations have a central icosahedral
cell like this,  
but it is also found in noncentric parts of icosahedral clusters with a sufficiently large 
anti-Mackay overlayer. The elongated pentagonal pyramid shown in Fig.~\ref{fig:shells}(b) possesses only 
25 contacts in the shell and is found in structures with icosahedral or decahedral symmetry. 
The cuboctahedron depicted in Fig.~\ref{fig:shells}(c) is an example for a cell with 24 shell contacts and 
is the seed of compact fcc geometries but also occurs in larger structures, 
since any LJ icosahedron can be considered as a combination of 20 fcc-tetrahedra~\cite{doye2}.
Near the surface, also incomplete icosahedra with 11 nearest neighbors  
forming 25 contacts are found [Fig.~\ref{fig:shells}(d)], 
which we will not distinguish from complete icosahedral cells in the following.

Based on these features, icosahedral cells in a given conformation are easily identified
in its contact map.
We consider two monomers $i,j$ as being in contact if $r_{ij}<0.8$ 
which is slightly larger than the optimal LJ distance $r_0$. 
Eventually, the total number of icosahedral cells $n_{\rm ic}$ is an appropriate parameter
to determine the overall geometry, at least at low temperatures: 
If $n_{\rm ic} >1$, the cluster is in an icosahedral state with a larger anti-Mackay overlayer, 
while for $n_{\rm ic}=1$ the shell is of Mackay type. 
For $n_{\rm ic}=0$, the cluster is non-icosahedral (e.g., fcc-like or 
decahedral). 

\begin{figure}
\centerline{\epsfxsize=8.8cm \epsfbox{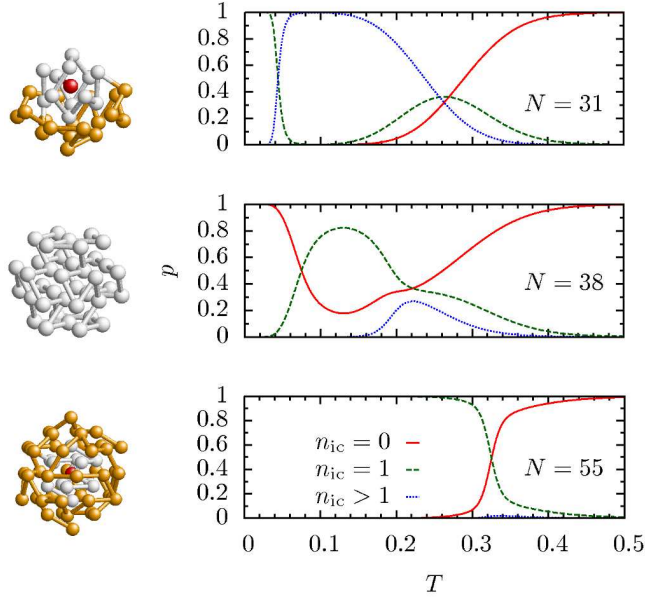}}
\caption{\label{fig:p_wpart31} 
Temperature dependence of the probability of 
icosahedral (Mackay: $n_{\rm ic}=1$, anti-Mackay: $n_{\rm ic}>1$) and \mbox{nonicosahedral} 
($n_{\rm ic}=0$) structures for exemplified polymers with lengths $N=31,38$, and $55$.
In the left panel, the corresponding lowest-energy morphologies are shown.
}
\end{figure}

In the following, thermodynamic properties of structure formation are analyzed for exemplified
polymers ($N=31,38,55$) with respect to the icosahedral content.
These examples are representatives for the different relevant structuring scenarios accompanying
the crystallization.
Figure~\ref{fig:p_wpart31} shows the respective populations $p$ for the three structural morphologies, 
parametrized by $n_{\rm ic}$, as a function of temperature. 

For $N=31$, we find that liquid structures with $n_{\rm ic}=0$ dominate above $T=0.5$, i.e.,
no icosahedral cells are present. Decreasing the temperature and passing $T\approx 0.4$, 
nucleation begins and the populations of structures with icosahedral cells ($n_{\rm ic}=1$ and 
$n_{\rm ic}>1$) increase. The associated energetic fluctuations are also signaled in the specific
heat [see inset of Fig.~\ref{fig:flucts} (a)]. 
Cooling further, $n_{\rm ic}$ is always larger than $1$, showing that the remaining monomers 
build up an anti-Mackay overlayer and create additional icosahedra. Very close to the temperature 
of the ``solid-solid'' transition near $T\approx 0.04$, where also the specific heat exhibits 
a significant peak, the transition from anti-Mackay to Mackay overlayer occurs
and the ensemble is dominated by frozen structures containing a single icosahedral cell surrounded by
an incomplete Mackay overlayer.

The exceptional case of the 38mer shows a significantly different behavior. A single icosahedral core
forms and in the interval $0.08<T<0.19$ icosahedra with Mackay overlayer are dominant.
Although the energetic fluctuations are weak, near $T\approx 0.08$, a surprisingly strong
structural crossover to nonicosahedral structures occurs: the formation of a maximally compact 
fcc truncated octahedron.
 
Finally, the ``magic'' 55mer exhibits a very pronounced transition
from unstructured globules to icosahedral conformations with complete Mackay overlayer
at a comparatively high temperature ($T\approx 0.33$). Below this temperature,
the ground-state structure has already formed and is sufficiently robust to resist the
thermal fluctuations.

\section{Concluding remarks}
In this study, we have precisely investigated thermodynamic properties of the liquid-solid 
transition of elastic flexible polymers of finite length and demonstrated that the 
formation of icosahedral and nonicosahedral structures proceeds along similar lines as for 
clusters of atoms governed by van der Waals forces. Morphologies of polymer ground-state 
conformations coincide with corresponding LJ clusters. However, covalent bonds between
adjacent monomers reduce at higher temperatures
the entropic freedom and thus the behavior of polymers in both the 
liquid and random-coil phase is different.
Introducing a structural ``order'' parameter probing the icosahedral content of a structure, we
have also analyzed the nucleation process in detail, which typically starts with the 
formation of an icosahedral seed, even for polymers preferring nonicosahedral lowest-energy
conformations.  
\section*{Acknowledgments}
This work is partially supported by DFG Grant  
Nos.\ JA 483/24-1/2, Leipzig Graduate School of Excellence ``BuildMoNa'', DFH-UFA PhD College CDFA-02-07, and 
NIC J\"ulich under Grant No. hlz11. 

\end{document}